\documentclass[apj]{emulateapj}

\usepackage{epstopdf}
\usepackage{amsmath}

\begin{document}

\title{Optimizing Doppler Surveys for Planet Yield}
\author{Michael Bottom\altaffilmark{1}, Philip S. Muirhead\altaffilmark{1}, John Asher Johnson\altaffilmark{1}, Cullen H. Blake\altaffilmark{2,3}}

\email{mbottom@astro.caltech.edu}
\altaffiltext{1}{Department of Astrophysics, California Institute of Technology, MC 249-17, Pasadena, CA 91125, USA}
\altaffiltext{2}{Department of Astrophysical Sciences, Princeton University, Peyton Hall, Ivy Lane, Princeton, NJ 08544, USA}
\altaffiltext{3}{Department of Physics and Astronomy, University of Pennsylvania, 209 S. 33rd St., Philadelphia, PA 19104}

\begin{abstract}
One of the most promising methods of discovering nearby, low-mass planets in the habitable zones of stars is the precision radial velocity technique. However, there are many challenges that must be overcome to efficiently detect low-amplitude Doppler signals.  This is both due to the required instrumental sensitivity and the limited amount of observing time.  In this paper, we examine statistical and instrumental effects on precision radial velocity detection of extrasolar planets, an approach by which we maximize the planet yield in a fixed amount of observing time available on a given telescope.  From this perspective, we show that G and K dwarfs observed at 400-600 nm are the best targets for surveys complete down to a given planet mass and out to a specified orbital period. Overall we find that M dwarfs observed at 700-800 nm are the best targets for habitable-zone planets, particularly when including the effects of systematic noise floors.  Also, we give  quantitative specifications of the instrumental stability necessary to achieve the required velocity precision. \end{abstract}


\section{\textbf{Introduction}}
Over 400 extrasolar planets have been discovered by surveys using the precision radial velocity (PRV) method \citep{wright2011}.  The first discoveries of planets using the PRV method were mainly massive planets in close-in orbits, known as hot Jupiters, revealing a surprising diversity in planetary systems \citep{mayorqueloz1995, butlermarcy1996}.  More recently, technical  advances have resulted in discoveries of planets with masses intermediate to terrestrial and gas giants, the ``Super Earths'' \citep{macarthur05, rivera2005, udry07, howard11, pepe11}
.  Future instrumentation promises the sensitivity needed to detect Earth analogs around Sun-like stars, but there are substantial challenges that must be overcome to attain this level of precision.

There are three kinds of limits to how well one can recover a radial velocity signal from a target star.  First, there are statistical limits that come from the signal-to-noise ratio and the depth, density, and shape of the spectral lines. These effects may be further separated as having components that come from the physical properties of the star, such as the stellar luminosity and distance to the star, which affects the signal-to-noise ratio of the spectrum; the chemical composition, which sets the number and depth of spectral lines; and the rotation rate, which affects the broadening of the spectral lines. Additionally, there are effects that come from observing, such as the exposure time, telescope diameter, system throughput (sky to detector), and detector noise, which affect the overall signal-to-noise ratio; the resolution of the spectrometer and the sampling of the line spread function, which affect the width and clarity of the observed lines; and the decision of which wavelength range to observe. This latter point is notable as instruments are optimized to observe in a particular region of the electromagnetic spectrum, and the line density and quality can vary significantly between various types of stars in different regions of their spectra.  Furthermore, for ground-based studies, there are regions where telluric absorption makes observing impossible.
       
Even if all the negative effects above are minimized, there is still an important second class of problems which will hurt the velocity precision: the inability to properly control and/or characterize the changes in the instrumental profile of the spectrometer from night to night. Guiding errors, such as a star moving on a slit, can be a major component of this, as displacements in the center of light lead to skews in the instrumental profile.  Additionally, small changes in the ambient pressure, temperature, etc. lead to changes in the optical path, which are degenerate with velocity shifts. This was realized as being a fundamental limit to velocity precision over 45 years ago \citep{griffin1967}, and successful attempts to mitigate this require control of the environmental conditions and also active modeling of the fluctuating instrumental profile.  

Finally, even in the limit of a perfect instrument and infinite observing time, there are wavelength-dependent stellar effects that can mimic the signal of a planet orbiting a star, such as starspots, stellar activity (ÒjitterÓ), and quasi-periodic oscillations of the stellar photosphere. These deleterious effects may be mitigated somewhat by clever observing strategies or modeling \citep{dumusque2011a,dumusque2011b}, and the most successful programs are sometimes limited by this class of problems.

Many studies, notably \citet{bouchy2001, reiners2010, wang2012} have considered the best ways to detect planets around stars.  However, we take a somewhat different approach, where our end goal is to optimize a radial velocity \textit{survey} for sensitivity and detection efficiency.  Optimizing an observing plan to detect a planet around a particular star has a different set of requirements than a plan that seeks to discover the maximum number of planets, or discover a particular class of planets.  When considering how to optimize a survey, all the effects  mentioned in the previous paragraphs must be considered together with the expected velocity signals caused by planets.  Less massive stars are dimmer, but have higher velocity signals from similarly-sized planets.  Additionally, there are more nearby low-mass stars than high-mass stars, given the present-day stellar mass function.  In order to select targets for a survey, a balance must be struck, which will depend critically on the wavelengths of observation.  As we will show, it will also depend on the particular kind of survey under consideration.

In this work, we will characterize the statistical limitations in radial velocity observations under the consideration of finite observing time.  We will consider the most productive way to choose targets for a survey in order to maximize the recovery of \textit{planets per unit of observing time}.  

We arrive at our conclusions by the following chain of reasoning: in Section 2, we examine the precision achievable on different types of stars at a fixed distance and with fixed observing time, assuming a perfect instrument.  In Section 3, we then relax the assumption of fixed observing time and discuss the time necessary to detect different types of planets.  We remove the assumption of a fixed stellar distance by considering nearby stars and the effects of the present-day mass function, and discuss the optimal way to select targets as a function of observing wavelengths.  In Section 4, we discuss the effects of instrumental noise floors, how they arise, and how they affect the results in the previous sections.

\section{Effects of Stellar type}

Previous studies have sought to characterize the fundamental radial-velocity quality of different stellar spectra, considering the number density and depth of spectral lines.  Additionally, these studies have examined the fundamental limitations set by the signal-to-noise of an observation (e.g. \citet{bouchy2001}).  These effects are not completely separate; they depend on the spectrum's shape and its specific intensity, modulated by the effect of stellar size and distance.  If velocity precision is the ultimate goal, this requires additional consideration of at least one other thing, that being the observing wavelengths.

We choose to examine the effects of spectrum and wavelength range on velocity precision. We begin by using the latest spectral models \citep{allard2011} based on the PHOENIX code \citep{hauschildt1999} in the T$_{\rm eff}$ range of 2600 to 6200K, [Fe/H]=0, from 0.3 to 2.5 $\mu$m. For stars of T$_{\rm eff}$ = 2600 - 3400 K, we choose models where $\log{g} $ = 5.0; for 3600-5800 K we use $\log $ g = 4.5, and for 6000-6200 K we use $\log $ g = 4.0.  In order to simulate the effects of instrument resolution, we convolve the models with a generic Gaussian instrumental profile equivalent to a resolution of $\lambda/\Delta \lambda$ = 75000 and a sampling of 3.0 pixels per resolution element.  Next, we rotationally broaden the spectra of stars with T$_{\rm eff} > 3600$K to match a disk-integrated $v\sin{i} = 2.0$ km/s.  For cooler stars, we use $v\sin{i}$ values taken from \citet{reiners2010} (see Table \ref{simtable} for the exact values).  This can be considered the ÒperfectÓ observation, with the only degradation of spectral quality set by the spectrometer resolution and rotational broadening. To consider the effects of finite observing time, we then inject Poisson noise at a level equivalent to one minute of observing time, and Gaussian read noise at the level of 5 electrons/pixel spread over 10 pixels in the cross dispersion direction \footnote{for the typical (high) signal-to-noise levels in precision radial velocity observations, read noise does not contribute significantly to the error budget, a fact that we verified here}.  Note that fainter stars will have a lower S/N ratio, as will wavelength bands away from the peak of the spectral energy distribution.  

Determining a radial velocity is equivalent to trying to recover the shift in a spectrum with respect to a reference spectrum, according to $\Delta \lambda/\lambda = v/c$.  Algorithms to do this include cross-correlation maximization, least-squares minimization, forward modeling, and various simplex algorithms.  For our simulated observations, the degradation of the spectrum with noise will introduce a spurious shift with respect to the convolved, non-noisy spectrum, which we recover using an algorithm where we fit the peak of the cross-correlation function with a polynomial.  We checked that this converges to zero velocity shift linearly  the signal-to-noise ratio approaches infinity, as expected. This precision is what would result from having a perfect reduction pipeline, regardless of the specifics of the calibration method.  

\begin{figure*}[!h]
\centering
\plotone{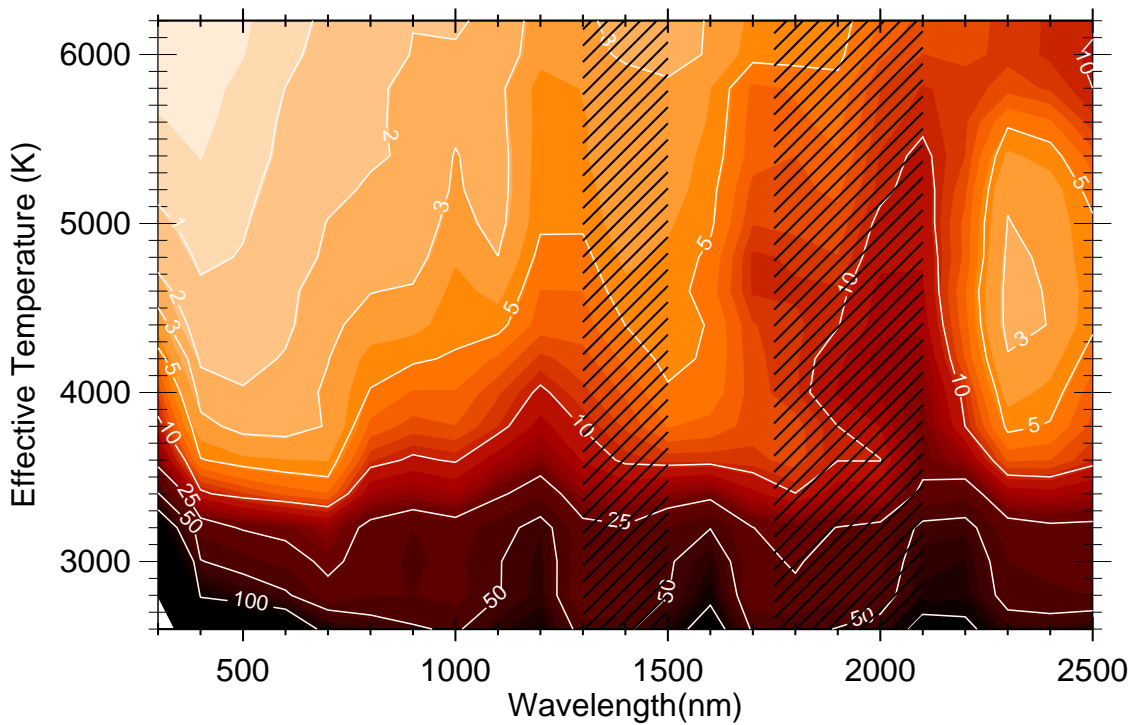}
\caption{Doppler precision as a function of wavelength range and star temperature for a \textit{fixed amount of observing time} (60 s).  The stellar spectra are derived from rotationally-broadened main-sequence templates from 2600-6200K, stepped in 200K increments, and the wavelength range is stepped in 100 nm increments.  The contours indicate the velocity precision in m/s.  From the perspective of velocity precision, the best result is achieved in the range of 400-600 nm.  The hashed regions correspond to wavelengths where the infrared absorption is too high for ground-based observations to be effective.  This simulation assumes a 1.28 m$^2$ telescope dish, a spectrograph with $R$=75000, and sky-to-detector throughput of 10\% (The full simulation parameters, including stellar parameters, are given in the Appendix).}
\label{vrmsteff}
\end{figure*}

As a measure of our velocity precision, we repeat this simulated observation three hundred times, and take the standard deviation of the radial velocities as a measure of the velocity precision, ($\sigma_v$).  In order to assess the best wavelengths of interest and target spectral classes, we repeat this simulation for wavelength regions of size 100 nm and stellar effective temperatures from 2600-6200 K at steps of 200 K, from 0.3--2.5 $\mu$m, with an assumed distance to the star of 10 pc.  The results are shown in Figure \ref{vrmsteff}.

The main conclusion of this numerical experiment is that in terms of best achievable \textit{radial velocity precision} in a fixed amount of observing time for a star at a fixed distance, there is little advantage to moving redward of 600 nm, and the best ÒoverallÓ area to observe for a large range of spectral classes is the region between about 400 and 600 nm.  This result is based on the convergence of a few physical effects.  First, the spectral energy distribution peaks in this wavelength range for stars in the temperature range 4750-6000 K.  This gives an advantage in terms of signal-to-noise.  Second, high spectral quality occurs at these wavelengths, caused by deep atomic absorption features from metals.\footnote{Coincidentally, this area happens to overlap the absorption lines of molecular iodine, a commonly used wavelength calibration source, which helps to explain the success of iodine cells in planet hunting}  While the cooler stars do share some of these features, their spectral energy distributions peak at redder wavelengths. This is evident in the contours moving rightward in the 2600-4250 K, 600-800 nm area.  Additionally, the overall decrease in intensity of the stellar spectra at longer wavelengths leads to a general decrease in precision from left to right.  Furthermore, the deep and rich molecular lines present in cooler stars partially offset the fact that the stars are smaller and dimmer, leading to bands of higher radial velocity precision from 1.4-1.6 $\mu$m and 2.3-2.6 $\mu$m, but the absolute level is still well below the precision from 400-600 nm.  Any increase in the exposure time will only change the absolute scaling of the velocity precision; it will not affect the relative precision between different spectral types and wavelength bands.

There are a number of caveats to this analysis.  In the infrared bandpass considered (longer than $\sim$650 nm), there are are significant absorption features by water and oxygen in the Earth's atmosphere, as well as OH emission lines, which make ground-based studies more challenging or even impossible.  We have shaded the areas where infrared absorption is too high for effective observing from the ground.  For the other regions, this simulation assumes that these features can be modeled and subtracted effectively and have no effect on the velocity precision.  At high resolution, it is possible to identify telluric features and remove them, and recent results suggest that atmospheric calibration can subtract these lines effectively (stable to 10 m/s over 6 year timescales), and can even use these features as stable wavelength references to 2 m/s over short timescales \citep{figueira2010a}.  A notable recent study that includes the effects of imperfect telluric subtraction can be found in \citet{wang2012}. Even so, in our result, the photon errors are a few times higher, and dominate the error budget, despite the assumption of perfect sky subtraction.

Another important point, not obvious from the plot, is that the high average rotation rate ($\sim$ 9 km/s) of late M-dwarfs imposes a severe penalty on the velocity precision obtainable on these stars, as the lines are significantly blended at these speeds.  Repeating this simulation with a fixed rotation rate of 2 km/s for these stars leads to about a factor of two higher in precision for the same amount of observing time.  This is a major penalty, as a photon limited observation takes four times as long to get twice the velocity precision.  Furthermore, the read noise is a proportionally larger part of the total signal at these lower photon counts.  

The fixed integration time can reasonably be considered too severe a restriction, particularly for the cooler stars and at longer wavelengths, since similarly sized planet can cause a much larger reflex velocity on a smaller star.  Furthermore, the simulation considered different types of stars at a fixed distance from the Earth, which ignores the realities of the present day mass function.  These are both fair points, and we consider the latter effect of detectability of planets in the next section.

\section{Maximizing radial velocity survey yields}

While it is clear that hotter stars at a fixed distance away are more amenable to high velocity precision, the goal of most radial velocity surveys is not solely high precision, but planet detection.  Target selection is important.  Therefore, there are two major revisions to the above analysis.  The first is that planets of a given mass and orbital distance will cause a larger reflex velocity in lower mass stars.  This means that less observing time will be needed to detect that planet, making lower mass stars more attractive targets, as their relative faintness is somewhat mitigated.  Equally important is the distribution of stellar masses as a function of distance from the Sun---ie, the probability that a star exists a certain distance from the Sun depends on how massive it is.  Closer stars make better targets because because of the higher incident flux, though less massive stars have lower luminosities.

We first quantify the above statements.  The reflex velocity on a star caused by an orbiting planet is given by\footnote{This expression can be derived quickly from conservation of momentum ($M_{*}v_{*} = M_{pl} v_{pl}$) and \textit{Kepler}'s second law ($P^2 \propto a^3$), then scaling to the reflex velocity of Sun caused by the Earth, 8.9 cm/s.}
\begin{align}
v_{*} = 8.9 \  \mathrm{cm/s} \times \Bigg(\frac{M_{pl}}{M_{\oplus}}\Bigg)  \Bigg(\frac{M_{*}}{M_{\odot}}\Bigg)^{-1}  \Bigg( \frac{a}{1 \mathrm{AU}}\Bigg)^{-1/2}
\end{align}

\noindent where $M_{pl}$ is the mass of the planet, $M_{*}$ is the mass of the star, and $a$ is the orbital distance of the planet (we consider only circular orbits for simplicity.  Also, $M_{pl}$ should always be considered as $M_{pl} \sin i$ to take into account inclination).

Secondly, we consider the amount of observing time per night it takes to detect a hypothetical planet which causes a given reflex velocity in its parent star.  We consider a ``detection'' to be equivalent to a measurement with velocity precision equal to the reflex velocity of the star.  A simpler way of saying this is that in the limit of many observations evenly spaced out over the planetary orbital phase, a single-point precision of 1 m/s is sufficient to detect a planet causing a 1 m/s modulation of its parent star's velocity. We confirmed this using white-noise simulations of planet-induced stellar RV variations detected using  both the Lomb-Scargle periodogram\citep{scargle1982} and a $\chi^2$ test to reject the null hypothesis of constant stellar RV.

 We repeat the calculation of Figure \ref{vrmsteff} for 10 minute's worth of observing time, for a star at 10 pc, from Figure \ref{vrmsteff}.  For the more general case, we have a scaling:

\begin{align}
t=600 \ \mathrm{s} \times \Bigg[ \frac{v_{\mathrm{10 min, 10 pc}}}{v_{*}(M_{pl}, M_{*}, a_{pl})} \times \frac{d}{\mathrm{10 \ pc}} \Bigg]^2
\end{align}

\noindent where $v_{*}$ is the reflex velocity of the star; a function of the planet mass, stellar mass, and orbital distance, given in the previous equation.  This expression can be derived in two steps.  First, recall that the velocity precision scales as the inverse of the signal-to-noise ratio, which is proportional to the square of the observation duration.  Of course, this scaling relation only applies in regimes where read noise is negligible, but this is true in the cases under consideration, where typical observing times are many minutes. Second, the signal-to-noise is equal to the square root of the number of photons, which scales as the inverse of the distance to the star ($\sqrt{N_{ph}} \propto d^{-1}$).  Putting these two together gives the above expression, which is exact within our framework.  This last result is interesting in of itself, because it is the inverse of \textit{planets per unit observing time}, which we can evaluate as a function of wavelength of observation and stellar type.  

Equation 2 depends on the variables $M_{*}, M_{pl}, a_{pl}$, which determine $v_{*}$; on $\Delta \lambda_{obs}$ and $T_{\rm eff}$, the bandwidth of observation and stellar effective temperature, which determine the velocity precision in ten minutes for a star at 10 pc; as well as the distance to the star, $d$.  To simplify things and remove one parameter, we convert freely between stellar mass and effective temperature using the BCAH 98 isochrones \citep{bcah98} at 2 Gyrs ([Fe/H]=0), which are reasonably accurate for physical properties of low-mass stars.  In this $T_{\rm eff}$ range, stellar properties are not very sensitive to evolution for the first few Gyrs on the main sequence.

In principle, all that is needed now is a complete list of stars of known spectral type and distance from the Sun (this gives $d$ and $M_{*}$ in the above expression), and the planet mass/orbital distance distribution as a function of spectral type.  Then we can calculate the amount of time needed to detect a putative planet, as a function of observing wavelength range.  We can then order the list in terms of observing time, which will show what wavelength range and spectral type maximizes the number of planet detections.  

Unfortunately, the planet mass/orbital distance distribution function is not known, and there is hardly a complete list of stars in the galaxy with known distances and spectral types.  Despite this, we can construct a reasonably accurate stellar census using the RECONS ``100 nearest stars'' sample (www.recons.org), which is complete out to $\sim$7 parsecs.  We populate the first 7 parsecs of our sample from RECONS.   In extrapolating outwards, we assume a constant stellar number density per unit volume and proceeding outward in spherical shells, drawing from an estimate of the present day stellar mass function \citep{reid2002}.  As a check of this method, we found that this reasonably reproduces the number of stars of each spectral type in the RECONS 10 pc sample.\footnote{The full RECONS 10 pc sample has not been released as of the submission of this paper, though they have released the number of stars of each spectral class in the 10 pc sample.}, with the deviation of our results at  the few percent level, consistent with the statistical variability of the stellar neighborhood.

We do not attempt to guess the planet mass/orbital distance distribution as a function of spectral type; after all, discovering this is one of the goals of planet surveys.  However, we can apply our analysis to cases where we can make reasonable assumptions.  First, we consider surveys that are complete down to a certain planetary mass and orbital period.  Second, we focus on searching for planets in stellar habitable zones.

\subsection{Surveys complete to limits in planet mass and orbital period}

Since RV surveys detect stellar accelerations, for a given stellar mass, the quantity $M_{pl}a^{-1/2}$ determines the necessary velocity precision.  We can set this to a constant and then determine the necessary time to reach this precision as a function of stellar mass and distance:

\begin{align}
t=600 \ \mathrm{s} \times \Bigg[ \frac{v_{\mathrm{10 min, 10 pc}}\times M_{*}}{(\mathrm{8.9 \ cm/s}) \ (M_{pl} a^{-1/2}= \mathrm{const})} \times \frac{d}{\mathrm{10 \ pc}}\Bigg]^2
\end{align}

\begin{figure*}[!h]
\centering
\plotone{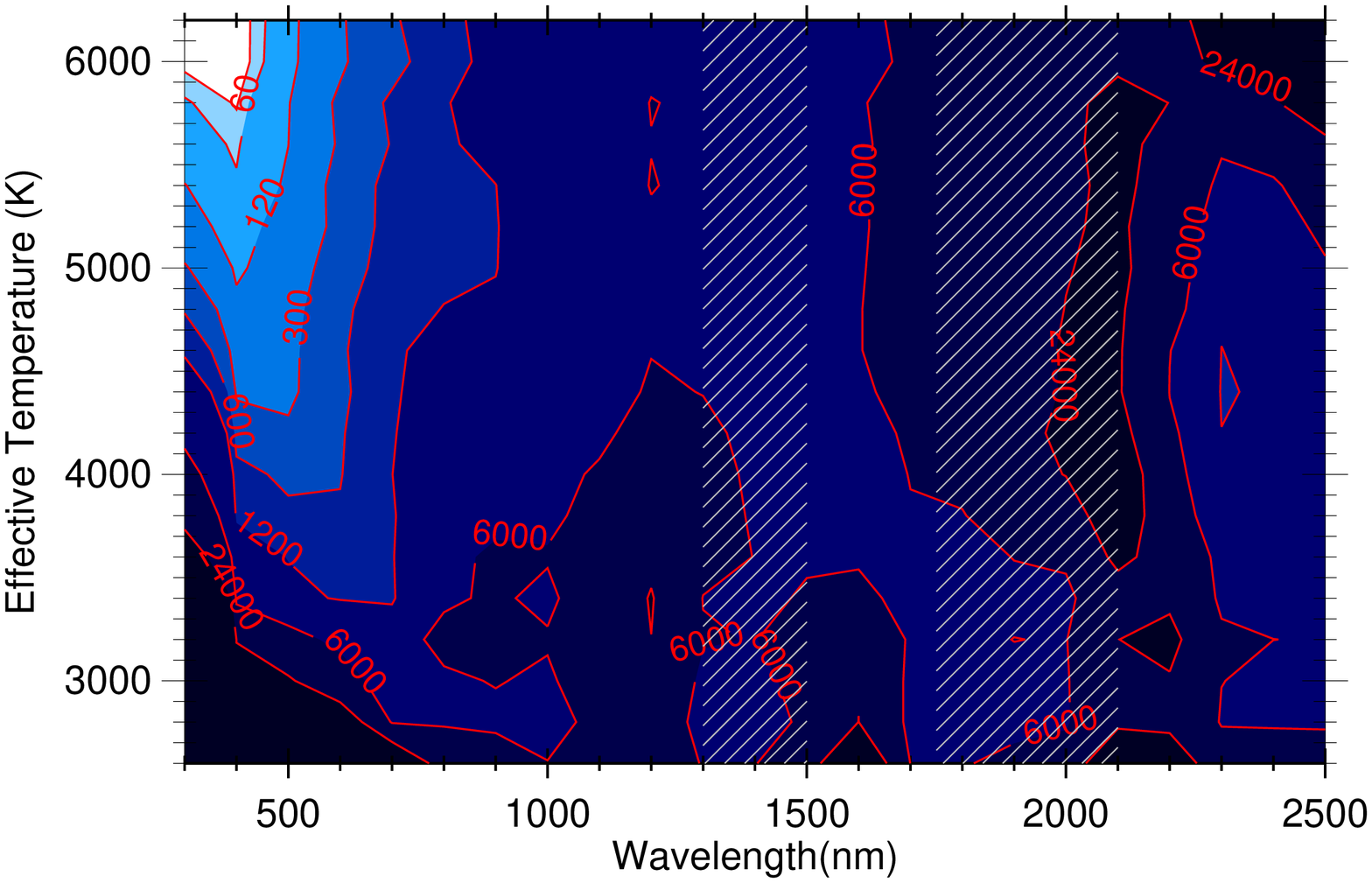}
\caption{The time (seconds) to detect ($\sigma_{v} = K$) a planet with $M_{pl} a_{pl}^{-1/2}= 5 M_{\oplus} (\mathrm{ 1AU})^{-1/2}$, 10 parsecs away, for a range of observing wavelengths and stellar effective temperatures.  The hashed regions correspond to wavelengths where the infrared absorption is too high for ground-based observations to be effective.  This simulation assumes a 1.28 m$^2$ telescope dish, a spectrograph with $R$=75000, and sky-to-detector throughput of 10\% (The full simulation parameters, including stellar parameters, are given in the Appendix).}
\label{time_fixed_mass_orbit}
\end{figure*}

A plot of this result with the distance fixed at 10 pc and $M_{pl}a^{-1/2} = 5 M_{\oplus} \mathrm{AU}^{-1/2}$ is shown in Figure \ref{time_fixed_mass_orbit}.  The higher mass stars generally take less time in terms of planet detection, and the region of 400-600 nm is still the best for a range of stars.

To include the effects of real stellar populations, we simulate performing such a survey.  We apply the results of Figure \ref{time_fixed_mass_orbit} to our simulated sample of stars out to 20 pc.  For each star, we calculate the amount of observing time necessary to get a radial velocity precision lower than the reflex velocity caused by the planet.  

\begin{figure*}[!h]
\centering
\epsscale{1.2}\plotone{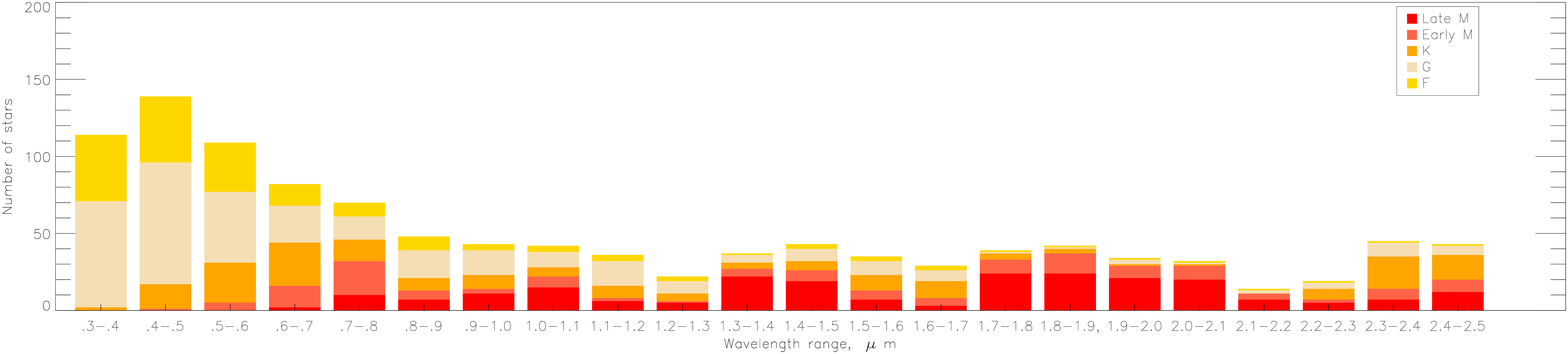}
\caption{The maximum number of observable stars each night, with the goal of achieving the velocity precision necessary for a detection of a planet with  $M_{pl} a_{pl}^{1/2}= 5 M_{\oplus} \mathrm{AU}^{1/2}$ on each star.   This is plotted as a function of observing wavelengths, assuming 9 hours of observing time per night and 2 minute acquisition time between targets.  The survey was simulated using the RECONS 7 pc sample and the present-day mass function \citep{reid2002}, and extends to 20 pc (going out to 300 pc makes little difference, as bright and nearby stars are the most time-efficient targets).  }\label{survey_fixed_mass_orbit}
\end{figure*}

In order to maximize the survey yield, we order these times in increasing order and select the stars until we exceed the amount of observing time per night (9 hrs, with 2 minutes of acquisition per target).  We plot this result in Figure \ref{survey_fixed_mass_orbit} as a function of wavelength range and spectral type (for our conversions from mass to spectral type, see the appendix).

The results in Figure \ref{survey_fixed_mass_orbit} show that the best place to observe is around 400-500 nm, with the primary targets being F, G, and K stars.  Redward of 600 nm, early M dwarfs become the primary targets, but the number of stars that can be observed from night to night is lower by a factor of two to four.  Repeating this experiment for a distance out to 300 pc gives roughly the same results, as nearby stars are the most efficient,  though there is a slight bias towards hotter stars in the visible wavelengths, because their luminosities make them observable over a larger volume.

The results here do not consider the effects of stellar activity, but it should be pointed out that F stars are known to be quite jittery and don't make ideal targets.  Regardless, they make up a relatively small percentage of the total stars, so the results do not change very much.

\subsection{Habitable zone planet surveys}

Next, we consider the case of planets in the habitable zones of their parent stars.  Here we can avoid the question of planet distribution as a function of orbital distance and spectral type.  For a star of a given T$_{\mathrm{eff}}$, we can calculate the inner and outer habitable zone, where we define the habitable zone to be the region where the equilibrium temperature of the planet is between 175 and 275 K \citep{kaltenegger2011}:

\begin{equation}
T_{eq} = T_{\mathrm{eff}} \left[ \frac{(1-A) R_{*}^2}{4 \beta a^2 (1-e^2)} \right]^{1/4}
\end{equation}

\noindent where we set the albedo, $A = 0.5$, the eccentricity $e=0$, and the planetary re-radiation fraction $\beta$ to 1 for non-tidally locked planets and 0.5 for tidally locked planets \citep{peale1977}.  We set the radius to be a function of the effective temperature of the star, and use values generated from the BCAH 98 isochrones \citep{bcah98}.  (For main sequence stars in this T$_{\mathrm{eff}}$ range, the physical properties of the stars (mass, radius, etc) are not greatly affected by stellar evolution over a reasonably long timescale, so we simply use the values from the 2 Gyr isochrones.)  With these assumptions in place, we can solve for the two values of $a$, the inner and outer habitable zones.  We then calculate the reflex velocity of the stars caused by these planets, given by

\begin{align}
v_{*} = 8.9  \ \mathrm{cm/s} \times \Bigg(\frac{M_{pl}}{M_{\oplus}}\Bigg)  \Bigg(\frac{M_{*}}{M_{\odot}}\Bigg)^{-1}  \Bigg( \frac{a_{HZ}}{1 \mathrm{AU}}\Bigg)^{-1/2}
\end{align}

\noindent once again using the BCAH 98 isochrones to derive a consistent mass estimate for the star from the effective temperatures, and placing the planet at the edge of the inner HZ.

We repeat the calculation of the observing time necessary to detect a 5 M$_{\oplus}$ planet in the center of the habitable zone of its parent star, given by

\begin{align}
t=600 \ \mathrm{s} \times \Bigg[ \frac{v_{\mathrm{10 min, 10 pc}}}{v_{*}(5 M_{\oplus}, M_{*}, a_{HZ})} \times \frac{d}{\mathrm{10 \ pc}} \Bigg]^2
\end{align}

\begin{figure*}[!h]
\centering
\plotone{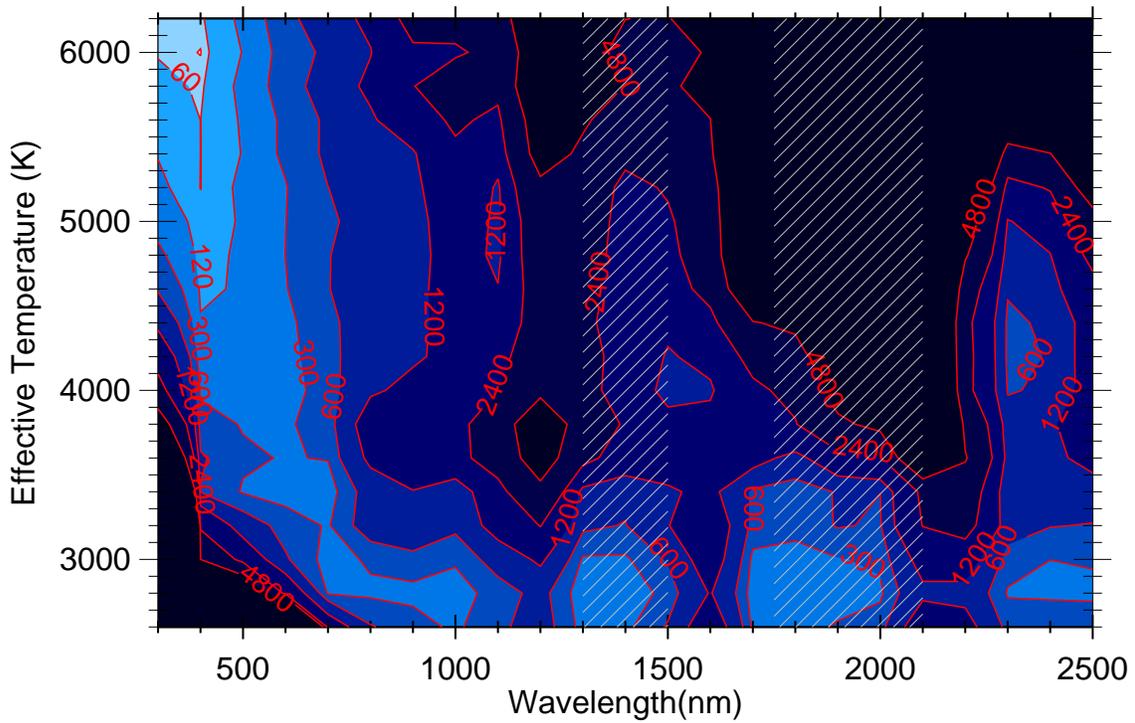}
\caption{The time (seconds) to detect ($\sigma_{v} = K$)  a 5 M$_{\oplus}$ planet in the habitable zone of its parent star, 10 parsecs away, for a range of observing wavelengths and stellar effective temperatures.  The hashed regions correspond to wavelengths where the infrared absorption is too high for ground-based observations to be effective.  This simulation assumes a 1.28 m$^2$ telescope dish, a spectrograph with $R$=75000, and sky-to-detector throughput of 10\% (The full simulation parameters, including stellar parameters, are given in the Appendix).}
\label{time_HZ}
\end{figure*}

This result gives a notion of \textit{habitable planets per unit observing time}, even if it depends on the distance to the star.  A plot of this result, for a fixed distance of 10 parsecs is shown in Figure \ref{time_HZ}.

Here, the M-dwarfs easily make the most attractive targets, as it takes much less observing time to recover a habitable zone planets around them.  Furthermore, this result ignored systematic noise floors, which will make some of the brighter stars problematic as targets for habitable zone planets, as their reflex velocities can be below the noise floor.

\begin{figure*}[!h]
\centering
\epsscale{1.2}\plotone{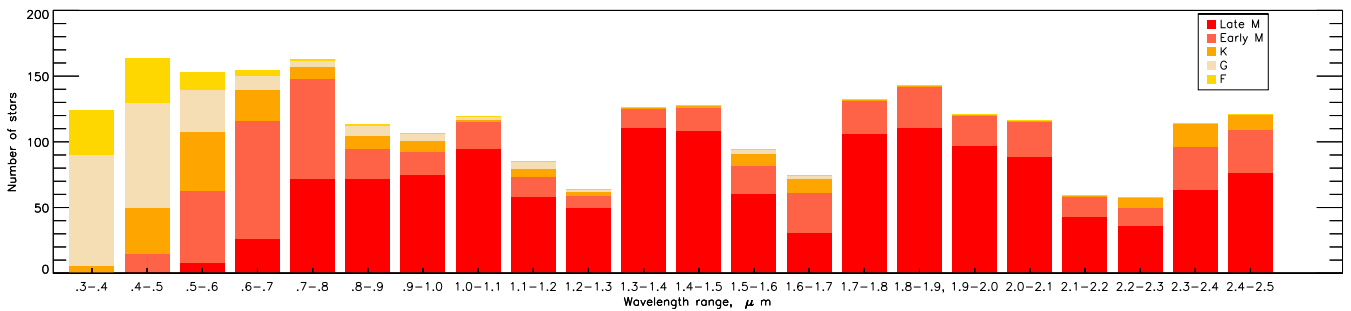}
\caption{The maximum number of observable stars each night, with the goal of achieving the velocity precision necessary for a detection of a 5 M$_{\oplus}$ habitable-zone planet on each star.  This is plotted as a function of observing wavelengths, assuming 9 hours of observing time per night and 2 minute acquisition time between targets.  The survey was simulated using the RECONS 7 pc sample and the present-day mass function \citep{reid2002}, and extends to 20 pc (going out to 300 pc makes little difference, as bright and nearby stars are the most time-efficient targets).  }
\label{survey_HZ}
\end{figure*}

Again, for the simulated stellar neighborhood of 20 pc, we calculate the observing time for each star necessary to detect a habitable zone planet according to the results of Figure \ref{time_HZ}.  We assume the same 9 hours of observing time per night at 2 minute acquisition time between targets, and plot the results in Figure \ref{survey_HZ}.  Here, there are dramatically different results compared to the previous case.  Because the habitable zone orbital distance decreases so rapidly for lower mass stars, the corresponding radial velocity signal is much larger for the same mass planet.  This advantage is so pronounced that late M dwarfs become the primary targets for surveys operating at essentially all wavelength ranges except the bluest.  The best wavelength to observe now becomes 700-800 nm, but this is somewhat constant for the range of 400-800 nm.  Note that the absolute number of targets are somewhat higher for the habitable zone survey.  This is an effect of the observing strategy; the observing time for each target is set by the \textit{required} velocity precision, not the \textit{ultimate} velocity precision of the instrument.  If one can detect a habitable-zone planet at 5 m/s in 1 minute, getting twice the velocity precision in four times the observing time is not worth it, even if the spectrometer is able to reach the precision easily.  Figure 5 may give the impression that there is little point to moving to the near-infrared, but this is somewhat an artifact of the choice of a 5 M$_{\oplus}$ planet as the target and the fact that systematic noise floors are not considered yet.  For a lower mass planets and more massive stars, the radial velocities of the stars are so small that they would almost certainly be below the systematic noise floor of the instrument, meaning that the \textit{most} promising targets would be in the near infrared, where the stellar radial velocities would be higher.  See Section 4.3 and Figure 9 for further discussion of these effects.

As previously mentioned, we did not consider the ``contamination'' of stellar spectra by telluric lines in the Earth's atmosphere, effects which are  wavelength-dependent.  In the case of the ideal survey complete to a mass-orbital distance limit, telluric lines are not relevant, because observations will be taking place in the visible wavelengths at 400-600 nm, where the atmosphere is mainly transparent.  For the habitable-zone survey, they will certainly be present in many of the infrared regions.  However, many telluric lines are stable at the the few m/s level, which is often \textit{lower} than the reflex velocities of the low mass stars, the primary targets of the survey.  This means telluric lines could be used as broadband wavelength references.  Due to the potential of M-dwarf infrared surveys, efforts put towards improving telluric referencing and calibration are highly important \citep{blake2011}.

Notably, we did not consider stellar jitter caused by oscillations, granulation, or activity such as spots.  For the interested reader, a thorough exploration of these subjects can be found in \citet{dumusque2011a, dumusque2011b} where these effects are analyzed in detail for solar-type stars, and observing strategies/corrections to mitigate different kinds of jitter are explored.  Of these three effects, the latter is considered the most troublesome in terms of planet detection, as the characteristic timescales of spot-related jitter are similar to planetary periods.  The simulations in this paper deal primarily with statistical errors and their dependence on observing wavelengths and stellar effective temperatures; the effects of stellar jitter are categorically different, as they are real radial velocity signals with the potential to confuse the actual planetary signal.  Furthermore, the spectral-type dependence of spot number, size, and shape are not well constrained, making meaningful simulation of these effects difficult within our framework.  It is possible that spot jitter would skew the results above; for example, if  one spectral type typically has spot distributions that are extremely stable and similar to planetary signals, whereas another has spots that are easily distinguishable from planetary signals, then the latter would be preferable to the former.  Of course,  this would be dependent on the ability of the data reduction and observing strategy to distinguish spot jitter from true planetary signals, whereas the results above show more fundamental limitations.  With this in mind, recent advances \citep{lanza2011, aigrain2012} have shown a promising ability to subtract out spot jitter with the combination of high-precision photometry.  Perhaps most encouragingly, by modeling stellar effects, Dumusque and collaborators were able to discover an Earth mass planet around $\alpha$ Centauri B, a star with a ``stellar noise'' level many times higher than the planetary signal \citep{2012Natur.491..207D}.

\section{Instrumental effects and systematic noise floors}

The most unrealistic assumption so far is the assumption of a perfect instrument in our hypothetical survey, as all past, present, and future instruments have a limiting precision.  In practice, this means that after a certain point, increasing the exposure time does not lead to an increase in velocity precision.  

Up to this point, we have assumed that we have a particular instrumental configuration; a resolution 75000 spectrometer with 3 pixel sampling and a velocity precision limited by the signal-to-noise ratio and read noise (which is negligible for most cases).  We first consider our choices of resolving power and sampling, and how they contribute to the velocity precision.  Furthermore, we consider the effects of instrumental instabilities and show how they can lead to systematic noise floors.  Finally, we assess the effects of these noise floors on our hypothetical survey, and show how they can substantially affect the optimal target selection.

\subsection{Effects of Spectrometer Resolution and Sampling}

The choice of resolving power of a spectrograph is important, as more sharply resolved lines lead to higher Doppler precision.  A simple analytic calculation for one line suggests that the accuracy should scale as $\sigma_v \propto$ R$^{-1}$, which was approximately reproduced in an early study  \citep{hatescochrane1992}. 

\begin{figure}[h]
\centering
\plotone{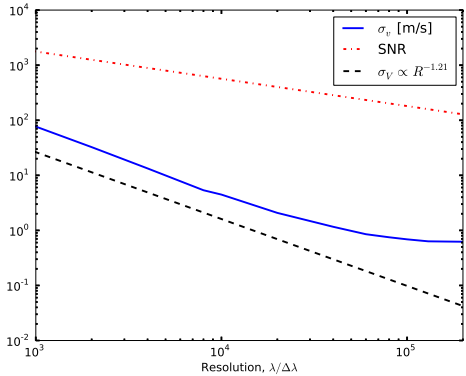}
\caption{The resulting velocity precision and signal-to-noise ratio for an observation of a Sun-like star with varying resolution, (3.0 samples/resolution element).  Note that the units on the vertical axis are m/s for the blue curve (velocity precision), and unitless for the red curve (signal-to-noise ratio).  The exposure time is held fixed, and the resulting signal-to-noise ratio \textit{decreases} as resolution increases, since less photons are incident per pixel.  Increasing the resolution always improves the velocity precision, but the point of diminishing returns is reached at about R = 45,000, which corresponds to the point where almost all the spectral lines are fully resolved.}
\label{res}
\end{figure}

We examine the effect of spectrometer resolution on velocity precision at a fixed exposure time on a Sun-like (Teff=5800 K, log g = 4.5 and [Fe/H]=0) star. We simulate one thousand measurements at resolving powers ranging from 10,000 to 150,000, in the wavelength range of 500-600 nm, and as before, take the standard deviation of the velocity measurements as our velocity precision.  The results are shown in Figure \ref{res}.

Our results show that the best-fitting power law is  $\sigma_v \propto$ R$^{-1.2}$, and that diminishing returns appear at about R=45000, where the spectral lines are resolved.  It should be mentioned that the power in the proportionality is a function of the spectral quality (for example, the line density), but we were unable to find much deviation from the value of 1.2 in interesting spectral regions.  We were not able to find any region where the power was as high as 1.5.  Additionally, although precision improves with higher resolution, resolving powers above 100,000 yield little additional benefit, as essentially all the lines are resolved.

Satisfying the Nyquist sampling theorem requires that at least 2 pixels cover each resolution element, but more pixels may be used.  Increasing the sampling turns out to have \textit{no effect on velocity precision, assuming the exposure time stays constant.}  The reason for this may be seen in that the number of photons per pixel is reduced by a factor of $N$, where $N$ is the number of pixels per resolution element; and the number of pixels in the data product is increased by the same factor.  Since the radial velocity precision is inversely proportional to the to signal-to-noise ratio ($\propto N^{1/2}$) and to the square root of the bandwidth, these effects cancel out.  We verified this (non)effect with numerical simulations.  We point out that in the limit of extreme (cm/s) velocity precision, it is not safe to ignore pixel topology effects, as the pixels may vary in efficiency over the center to the edge.  This can become a problem when the line-spread function is minimally sampled.

Taken together, these two results show that our choice of resolution 75000 and sampling of 3 pixels in our simulated instrument was reasonable, and did not affect the recovery of radial velocities adversely.

\
\subsection{Effects of incompletely recovered instrumental profiles}

In general, the measured output from a spectrograph is the ÒintrinsicÓ spectrum of the object convolved with the instrumental profile (IP) or line spread function (LSF) of the instrument:

$$
m(\lambda)=\int_{-\infty}^{\infty} s(\lambda '){\rm IP}(\lambda - \lambda ') d\lambda'
$$

In this equation, m($\lambda$) is the measured spectrum, IP($\lambda$) is the instrumental profile (which is normalized to unity by conservation of flux), and s($\lambda$) is the ``true'' stellar spectrum.  The instrumental profile is fiducially a gaussian with a full-width-half-max equal to the resolution of the spectrograph.

A typical extraction of a radial velocity datapoint involves consideration of the entire spectral region.  First, the instrumental profile is extracted from a wavelength reference source, and after the observation, this instrumental profile is deconvolved from the stellar spectrum.  In cases of simultaneous calibration (as with an iodine cell), the full transmission spectrum is modeled.  Finally, the radial velocity datapoint is extracted from the shift in the spectrum with respect to the wavelength solution.

Properly characterizing the IP of the spectrometer is a challenging task, but is essential to recovering radial velocities accurately \citep{valenti1995} ), especially as all further steps depend on its characterization.  It is a function of the optical path, and hence depends on environmental parameters like temperature and pressure, as well as slit illumination and focus.  These parameters can change during and between observations, and thus the IP must be recalculated for each observation; that is, the IP varies with time.  Furthermore, since optical elements have wavelength-dependent properties, the IP is wavelength-dependent.  This means that in practice it is necessary to model an IP varying with wavelength, rather than a constant one.  It is clear that any change in the IP that can be accounted for and modeled is not relevant.  However, any changes not accounted for will be interpreted as radial velocity shifts--this is due to having two different IPs; the physical instrument profile, and the approximation that is deconvolved from the observation.

To model this, we convolve our model spectrum with a perturbed IP.  While it is obviously not feasible to examine every possible perturbation, we can derive useful Òrules of thumbÓ from characterizing simple cases.  We begin by restricting ourselves to a gaussian LSF with equivalent resolution of 75000 and 3.0 samples/resolution element as our ideal IP. We simulate the effects of LSF mismatch by convolving with a stellar spectrum (Teff=5800 K, log g=4.5 [Fe/H]= 0) with the ÒperturbedÓ spread function and trying to recover the velocity shift (which should be zero) with respect to the spectrum convolved with the ideal IP. As before, we repeat this many times and take the standard deviation of our derived velocities as our velocity precision.

In the first case, we consider consider a gaussian LSF with some skew added. Skewness ($\hat{\gamma_3}$) is measure of asymmetry of the distribution, and is a property that can be straightforwardly calculated for a particular instrumental profile. Physically, skewness in the IP results when a source moves perpendicular to the slit direction.\footnote{  A real example of this effect can be found in the \textit{Herschel} \citep{pilbratt2010} observer's manual. \ \ \ \textit{http://herschel.esac.esa.int/Docs/PACS/html/ch04s07.html, particularly figure 4.17}.}

 Mathematically, the skewness of a function is given by:

\begin{equation}
\hat{\gamma_3} = E \left[  \left( \frac{X-\mu}{\sigma} \right)^{3} \right]
\end{equation}

\noindent This equation is general and applies to any distribution.  For our purposes, we consider the skew-normal distribution with parameter $\alpha$, given by the function
\begin{equation}
f(x)=2 \phi(x) \Phi(\alpha x)
\end{equation}
where
\begin{equation}
\phi(x) = \frac{1}{\sqrt{2 \pi}}e^{-x^2/2}, \ \ \ \ \Phi(x) = \int_{-\infty}^{x}\phi(t)dt
\end{equation}
Note that the skew parameter $\alpha$ is not actually equal to the skewness of the distribution, which is a complicated function of $\alpha$:

\begin{equation}
\frac{\alpha}{\sqrt{1+\alpha^2}}= \sqrt{\frac{\pi}{2} \frac{  |\hat{\gamma}_3|^{\frac{2}{3}}  }{|\hat{\gamma}_3|^{\frac{2}{3}}+((4-\pi)/2)^\frac{2}{3}}}
\end{equation}
where the skewness ($\hat{\gamma}_3$) is recovered by inverting the equation above for $\alpha$.  Despite the rather opaque equation above, the usefulness of this parametrization is that it is, in a sense, a ``simple'' way of adding skewness in a distribution, and for $\alpha = 0$ one recovers 
the normal distribution, which is our fiducial way of representing an instrumental profile.

\begin{figure}[h]
\centering
\plotone{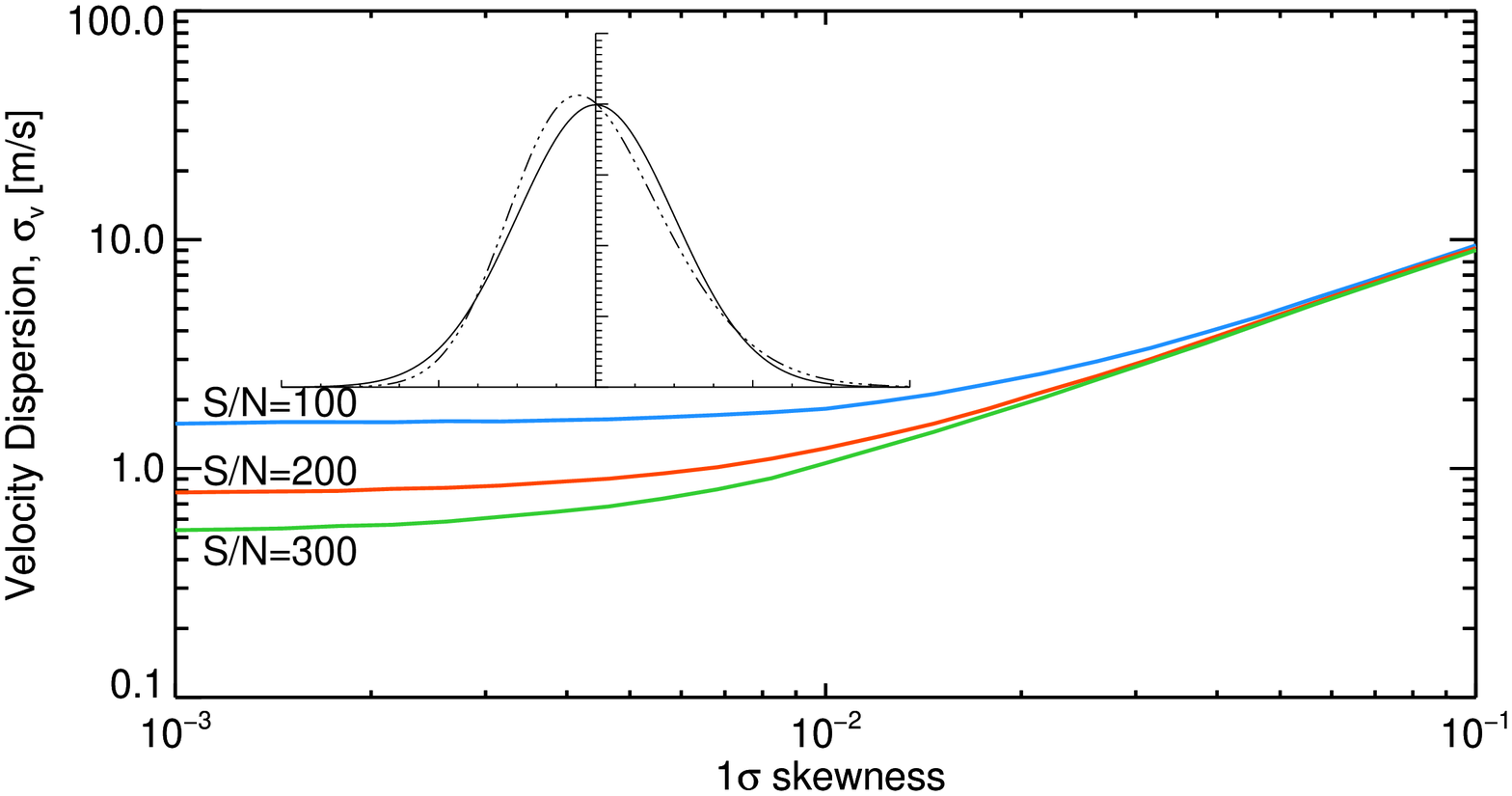}
\caption{The resulting velocity precision for observation of a Sun-like star with R=75000 (3.0 samples/resolution element), and skewness varying from 10$^{-1}$ to 10$^{-4}$. For reference, the inset shows a gaussian with skewness (not $\alpha$) of 0.3, an order of magnitude higher than the maximum value considered (none of the skew-normal distributions simulated have skews large enough to be visually distinct from a normal distribution.)  The skewness sets a signal-to-noise floor when it is greater than a part in 100, weakly dependent on signal to noise.  The flattening out of the curves occurs where the signal-to-noise ratio limits the velocity precision.}
\label{skew}
\end{figure}
                   
\begin{figure}[h]
\centering
\plotone{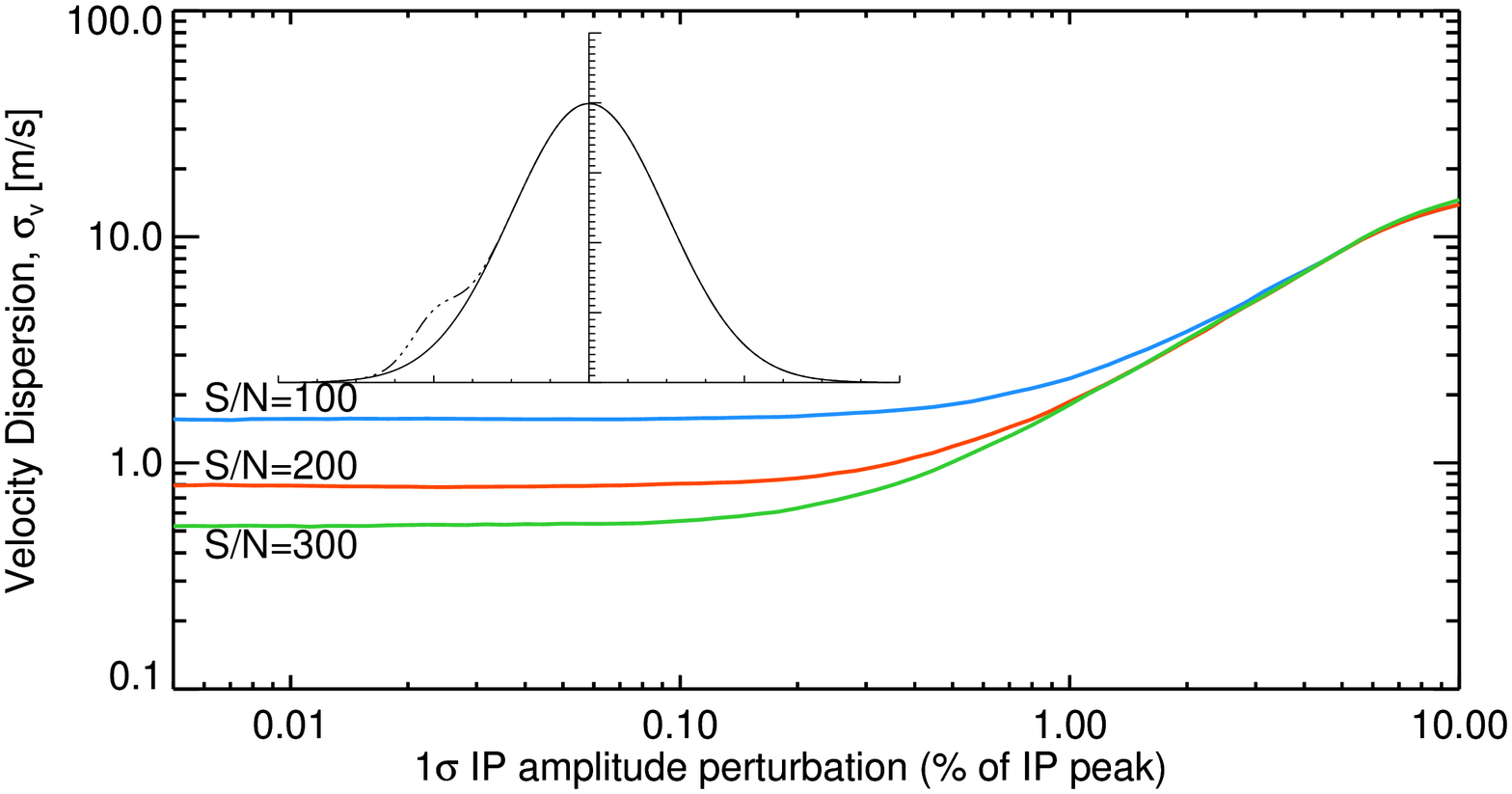}
\caption{The resulting velocity precision for observation of a Sun-like star with R=75000 (3.0 samples/resolution element), and perturbation amplitude varying from 10$^{-4}$ to 10$^{-1}$ of the peak amplitude of the LSF. For comparison, the inset shows a gaussian with a perturbation of 10\%, equal to the maximum value considered.  It is clear that for sub meter/sec precision, it is important that the perturbation amplitude of the distribution does not exceed 0.1 \%, a value weakly dependent on the signal-to-noise ratio.  The flattening out of the curves occurs where the signal-to-noise ratio limits the velocity precision.\vspace{2 mm}}
\label{pert}
\end{figure}

It is apparent from Figure \ref{skew} how a velocity floor can arise from an uncorrected skewed IP.  For example, at a skewness of 0.05, it will be impossible to do better than 5 m/s in precision, regardless of the signal-to-noise ratio.  

In the second case, we consider an IP with a small gaussian perturbation with varying amplitude. The choice of this form of perturbation is due to the fact that a common practical way to represent an imperfect IP is through many small gaussian functions added together on top of the main gaussian IP \citep{butlermarcy1996, endl2000, kambe2002}. The position of the perturbation is set to vary normally with a standard deviation equal to the standard deviation of the ideal IP (1). Also, the width of the perturbation is fixed to be of a characteristic size of one of the mini-gaussians used to model the IP.  We examine the effects of the perturbation amplitude on the velocity precision under the same conditions as the previous test.

The result of this simulation, in Figure \ref{pert}, reinforces how little tolerance there is in terms of characterizing the instrumental profile.  A perturbation as small as 3 \% IP peak is able to set a velocity floor of 5 m/s, independent of the signal-to-noise ratio.

\subsection{Effects of noise floors on survey yields}

The preceding section demonstrates the need of a stiff combination of stabilizing the IP through temperature, pressure, and illumination control and immediately capturing any changes that occur.  However, it is reasonable to assume that there will be some velocity floor in every survey.  We examined the effect by putting arbitrary noise floors at different velocities.  We repeated our simulated habitable-zone survey, except removed stars from the target list if their predicted planetary signal was below the noise floor (for the other survey, the very existence of noise floors negates its completeness).  This makes massive stars become inaccessible targets, and they are progressively replaced by the next less massive stars as the noise floor increases, and observations go to the limiting precision.  There are less targets per night as well, though the decrease is not as severe as one would think, as removing stars frees up available observing time for other targets within the detection limits:  G stars replace F stars, K stars replace G stars, and so on.

We simulated this effects for noise floors of 0.5, 1, 3, and 5 m/s.  The results are shown in Figure \ref{survey_vfloor5}, which demonstrates that mid and late-M dwarfs become better targets as precision decreases, and that the number of stars observable per night does not decrease substantially--after about 800 nm, the amount of observable stars in a given night is set primarily by the duty cycle.  Additionally, the wavelength range of 700-800 nm is the best overall in terms of total number of targets.  
\begin{figure*}[!h]
\centering
\epsscale{1.2}\plotone{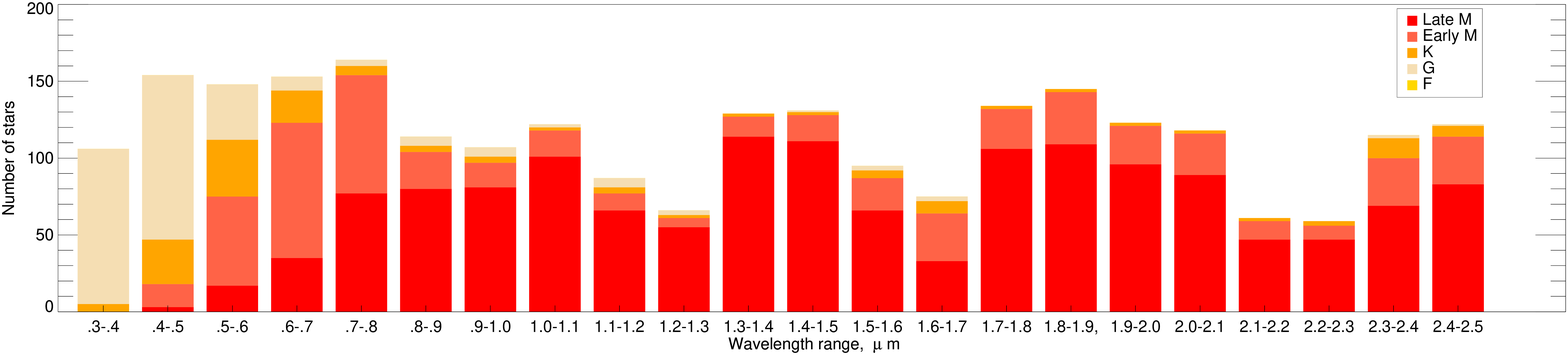}
\label{survey_vfloorp5}
\end{figure*}

\begin{figure*}[!h]
\centering
\epsscale{1.2}\plotone{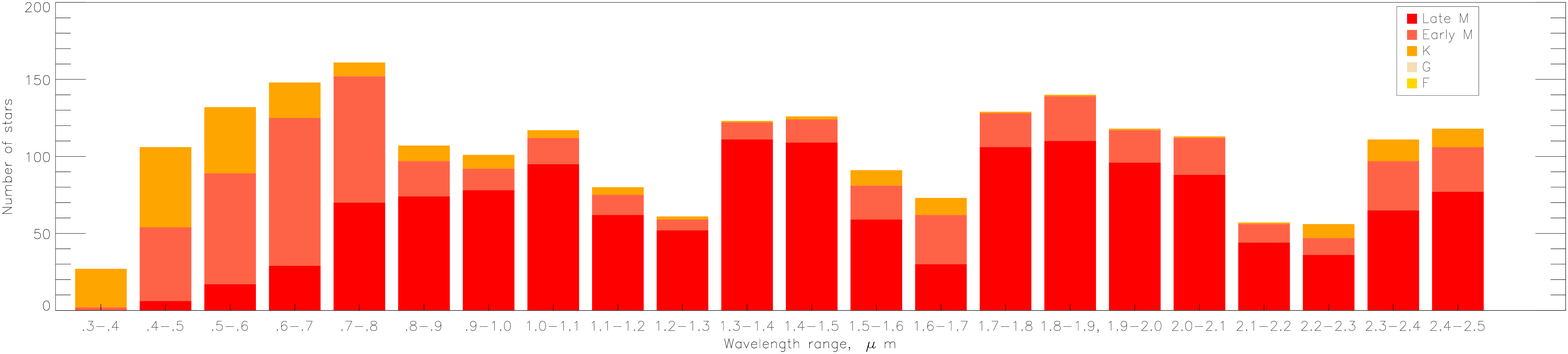}
\label{survey_vfloor1}
\end{figure*}

\begin{figure*}[!h]
\centering
\epsscale{1.2}\plotone{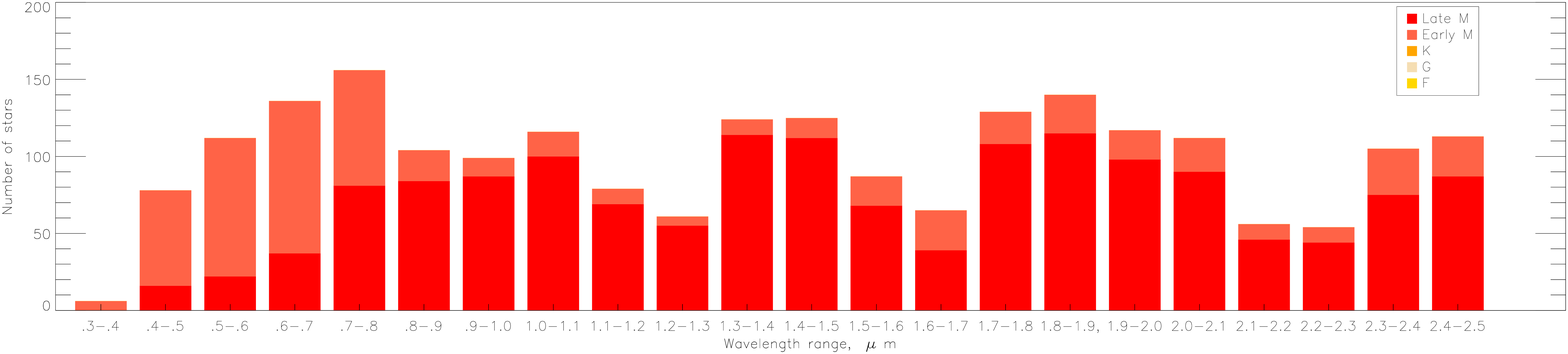}
\label{survey_vfloor3}
\end{figure*}

\begin{figure*}[!h]
\centering
\epsscale{1.2}\plotone{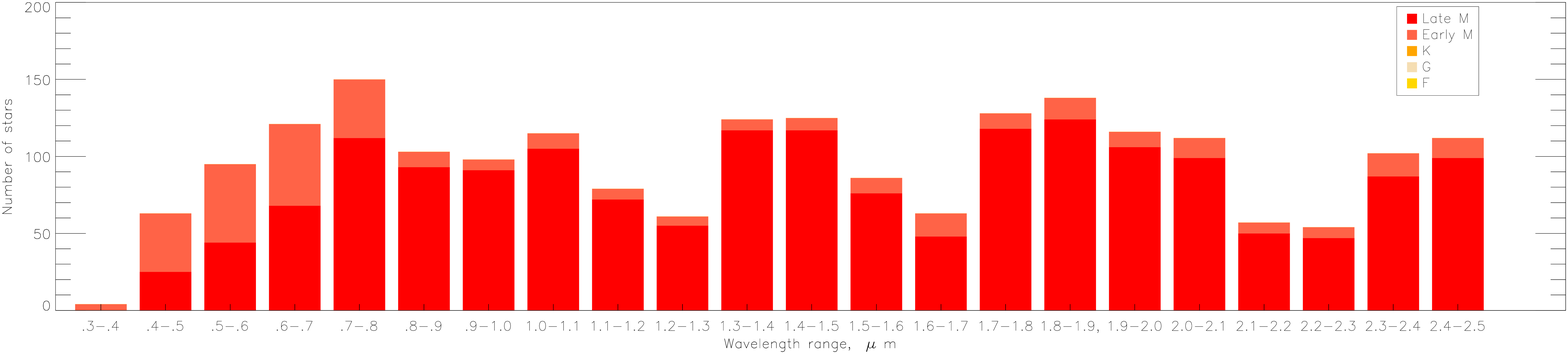}
\caption{The maximum number of observable stars as a function of observing wavelengths, assuming 9 hours of observing time per night and 2 minute acquisition time between targets. The survey was simulated using the RECONS 7 pc sample and the present-day mass function \citep{reid2002}, and extends to 20 pc (going out to 300 pc makes little difference, as bright and nearby stars are the most time-efficient targets).  These graph assumes a survey targeting habitable zone planets, with velocity precisions limited to 0.5 (top), 1, 3, and 5 m/s.  Furthermore, the region of 0.7-0.8 $\mu$ m is the best area to observe overall.}
\label{survey_vfloor5}
\end{figure*}

\section{Conclusion/Discussion}

We have investigated the design requirements and performance expectations for radial velocity surveys, deriving the best wavelength bands and targets for different survey goals.  For surveys targeting completeness out to a particular planetary mass-orbital distance product, the best targets are F, G, and K dwarfs observed at wavelengths spanning 400-600 nm.  For surveys of habitable-zone planets, the best targets are late M dwarfs in the wavelength range of 700-800 nm, though the number of possible targets stays flat from 400-800 nm.  Of the two survey methods, those searching for habitable-zone planets are more productive, as the larger expected radial-velocity signals lead to more targets, and hence more detections.  These results are based on the consideration of a number of input parameters, within the framework of maximizing planet detections in a fixed amount of observing time per night.  In particular, we considered how the spectral quality, stellar flux, photon noise, stellar and planetary mass, and stellar mass distribution in the galaxy play against each other to produce different ideal targets.

For surveys that are complete to a constant $M_{pl} a^{-1/2}$, brightness of targets is paramount: the best targets are hotter stars, and the visible wavelengths where these stars' spectral energy distributions peak, coincident with where the density of absorption lines is highest.  The target selection result holds even when considering the present-day stellar mass function.  Generally, the observing strategy consistent with this analysis is to target as many bright, non-jittery stars as possible out to a limiting magnitude (or volume), then move to lower masses.

In the case of habitable zone planet surveys, our results demonstrate the potential of infrared surveys of M dwarfs.  For an observing program that has relatively modest velocity precision, it is possible to have a survey of M dwarfs that is complete within the limits of the habitable zone.  Note that this requires a different sort of observing strategy than one where each observation is taken to the noise floor of the velocity precision.  This would make sense for F, G, and K stars, as the reflex velocities for habitable-zone candidates are very low.  However, for M-dwarfs, this is inefficient, as habitable-zone velocity precisions are higher than systematic instrumental limitations.  Since the velocity precision scales approximately as the \textit{square} of the observing time, getting excessive precision will dominate the nightly time budget.  This directly penalizes the number of target stars in the limit of complete phase coverage.  A better choice is to observe only until the \textit{target} precision is reached, and then move on to the next target.

We demonstrated how instrumental imperfections can lead to systematic noise floors, gave quantitative prescriptions for the level of stability needed in a restricted number of cases, and explored how these effects would change the scope of a habitable-zone planet survey.  Significantly, the number of potential targets is highly dependent on the noise floor for shorter wavelengths, but is basically unchanged for longer wavelengths.  The main change is that the target stars become later and later, demonstrating the high potential of infrared surveys.

\acknowledgements
The authors would like to thank (in alphabetical order) Prof. Lynne Hillenbrand (CIT), for providing guidance and expertise; Dr. Sasha Hinkley and Sebastian Pineda (CIT), for helpful critique and discussion; Dr. Peter Plavchan (IPAC), for detailed comments and perspective on early drafts of the paper.  We especially thank the anonymous referee, whose suggestions improved the paper substantially.

\clearpage
\appendix{}
\subsection{Simulation Parameters}
The following table gives the parameters in our simulation of velocity precision in Figure \ref{vrmsteff}.  It is stated in the text when any of these parameters are changed.

\vspace{3 mm}
\begin{center}
\begin{tabular}{llr}

\multicolumn{3}{c}{Simulation parameters} \\
\hline
Property 			& 	Default Value 	& Unit  \\
\hline
Stellar atmosphere models  & 	PHOENIX	 & BT-Settl 2009 \\
 				& [Fe/H]		& 0.0\\
				& $\log$ g \\
				& \ \ \ 2600 K $ \leq T_{\rm eff} \leq$ 3400 K		&  5.0\\
				& \ \ \ 3600 K$ < T_{\rm eff} \leq$ 5800 K		&  4.5\\
				& \ \ \ 5800 K $ < T_{\rm eff}$		&  4.0\\
				& $\alpha$-enhancement & 0.0\\
Stellar isochrone models & BCAH (1998) & 2 Gyr\\

Stellar rotation rate, $v \sin i$ \\
\ \ \ 2600 K$ \leq T_{\rm eff} < $ 2800 K	& 9.0 & km/s\\
\ \ \ 2800 K$ \leq T_{\rm eff} < $ 3200 K	       & 6.0 & km/s\\
\ \ \ 3200 K$ \leq T_{\rm eff} < $ 3800	 K	& 3.0 & km/s\\
\ \ \ 3800 K$ \leq T_{\rm eff}$				& 2.0 & km/s\\
Distance & 10 & pc\\
\hline
Telescope area & 1.28 & m$^2$\\
Observing time & 60 & s\\
Spectrograph resolution & 75000 & \\
Spectrograph sampling & 3.0 & pixels per resolution element\\
Throughput (sky to detector) & 10 & \%\\
Read noise & 5 & electrons\\
Cross-dispersion & 10 & pixels\\

\hline
\label{simtable}
\end{tabular}

\end{center}

\subsection{Spectral class conversions}

\begin{center}
\begin{tabular}{lc}

\multicolumn{2}{c}{} \\
\cline{1-2}
Spectral Class & Mass (M$_{\odot}$)\\
\hline
Late M  & 0.08--0.23\\
Early M   & 0.23--0.51 \\
K   & 0.51--0.79 \\
G & 0.79--1.05 \\
F & 1.05--1.6\\
\hline
\end{tabular}
\end{center}

\bibliographystyle{apj}
\bibliography{ms_bibliography}

\begin{thebibliography}{31}
\expandafter\ifx\csname natexlab\endcsname\relax\def\natexlab#1{#1}\fi

\bibitem[{{Aigrain} {et~al.}(2012){Aigrain}, {Pont}, \& {Zucker}}]{aigrain2012}
{Aigrain}, S., {Pont}, F., \& {Zucker}, S. 2012, \mnras, 419, 3147

\bibitem[{{Allard} {et~al.}(2011){Allard}, {Homeier}, \&
  {Freytag}}]{allard2011}
{Allard}, F., {Homeier}, D., \& {Freytag}, B. 2011, in Astronomical Society of
  the Pacific Conference Series, Vol. 448, 16th Cambridge Workshop on Cool
  Stars, Stellar Systems, and the Sun, ed. C.~{Johns-Krull}, M.~K. {Browning},
  \& A.~A. {West}, 91

\bibitem[{{Baraffe} {et~al.}(1998){Baraffe}, {Chabrier}, {Allard}, \&
  {Hauschildt}}]{bcah98}
{Baraffe}, I., {Chabrier}, G., {Allard}, F., \& {Hauschildt}, P.~H. 1998, \aap,
  337, 403

\bibitem[{{Blake} \& {Shaw}(2011)}]{blake2011}
{Blake}, C.~H., \& {Shaw}, M.~M. 2011, \pasp, 123, 1302

\bibitem[{{Bouchy} {et~al.}(2001){Bouchy}, {Pepe}, \& {Queloz}}]{bouchy2001}
{Bouchy}, F., {Pepe}, F., \& {Queloz}, D. 2001, \aap, 374, 733

\bibitem[{{Butler} \& {Marcy}(1996)}]{butlermarcy1996}
{Butler}, R.~P., \& {Marcy}, G.~W. 1996, \apjl, 464, L153

\bibitem[{{Dumusque} {et~al.}(2011{\natexlab{a}}){Dumusque}, {Santos}, {Udry},
  {Lovis}, \& {Bonfils}}]{dumusque2011b}
{Dumusque}, X., {Santos}, N.~C., {Udry}, S., {Lovis}, C., \& {Bonfils}, X.
  2011{\natexlab{a}}, \aap, 527, A82

\bibitem[{{Dumusque} {et~al.}(2011{\natexlab{b}}){Dumusque}, {Udry}, {Lovis},
  {Santos}, \& {Monteiro}}]{dumusque2011a}
{Dumusque}, X., {Udry}, S., {Lovis}, C., {Santos}, N.~C., \& {Monteiro},
  M.~J.~P.~F.~G. 2011{\natexlab{b}}, \aap, 525, A140

\bibitem[{{Dumusque} {et~al.}(2012){Dumusque}, {Pepe}, {Lovis},
  {S{\'e}gransan}, {Sahlmann}, {Benz}, {Bouchy}, {Mayor}, {Queloz}, {Santos},
  \& {Udry}}]{2012Natur.491..207D}
{Dumusque}, X., {et~al.} 2012, \nat, 491, 207

\bibitem[{{Endl} {et~al.}(2000){Endl}, {K{\"u}rster}, \& {Els}}]{endl2000}
{Endl}, M., {K{\"u}rster}, M., \& {Els}, S. 2000, \aap, 362, 585

\bibitem[{{Figueira} {et~al.}(2010){Figueira}, {Pepe}, {Lovis}, \&
  {Mayor}}]{figueira2010a}
{Figueira}, P., {Pepe}, F., {Lovis}, C., \& {Mayor}, M. 2010, \aap, 515, A106

\bibitem[{{Griffin}(1967)}]{griffin1967}
{Griffin}, R.~F. 1967, \apj, 148, 465

\bibitem[{{Hatzes} \& {Cochran}(1992)}]{hatescochrane1992}
{Hatzes}, A.~P., \& {Cochran}, W.~D. 1992, in European Southern Observatory
  Conference and Workshop Proceedings, Vol.~40, European Southern Observatory
  Conference and Workshop Proceedings, ed. M.-H. {Ulrich}, 275

\bibitem[{{Hauschildt} {et~al.}(1999){Hauschildt}, {Allard}, \&
  {Baron}}]{hauschildt1999}
{Hauschildt}, P.~H., {Allard}, F., \& {Baron}, E. 1999, \apj, 512, 377

\bibitem[{{Howard} {et~al.}(2011){Howard}, {Johnson}, {Marcy}, {Fischer},
  {Wright}, {Henry}, {Isaacson}, {Valenti}, {Anderson}, \&
  {Piskunov}}]{howard11}
{Howard}, A.~W., {et~al.} 2011, \apj, 726, 73

\bibitem[{{Kaltenegger} \& {Sasselov}(2011)}]{kaltenegger2011}
{Kaltenegger}, L., \& {Sasselov}, D. 2011, \apjl, 736, L25

\bibitem[{{Kambe} {et~al.}(2002){Kambe}, {Sato}, {Takeda}, {Ando}, {Noguchi},
  {Aoki}, {Izumiura}, {Wada}, {Masuda}, {Okada}, {Shimizu}, {Watanabe},
  {Yoshida}, {Honda}, \& {Kawanomoto}}]{kambe2002}
{Kambe}, E., {et~al.} 2002, \pasj, 54, 865

\bibitem[{{Lanza} {et~al.}(2011){Lanza}, {Boisse}, {Bouchy}, {Bonomo}, \&
  {Moutou}}]{lanza2011}
{Lanza}, A.~F., {Boisse}, I., {Bouchy}, F., {Bonomo}, A.~S., \& {Moutou}, C.
  2011, \aap, 533, A44

\bibitem[{{Mayor} \& {Queloz}(1995)}]{mayorqueloz1995}
{Mayor}, M., \& {Queloz}, D. 1995, \nat, 378, 355

\bibitem[{{McArthur} {et~al.}(2004){McArthur}, {Endl}, {Cochran}, {Benedict},
  {Fischer}, {Marcy}, {Butler}, {Naef}, {Mayor}, {Queloz}, {Udry}, \&
  {Harrison}}]{macarthur05}
{McArthur}, B.~E., {et~al.} 2004, \apjl, 614, L81

\bibitem[{{Peale}(1977)}]{peale1977}
{Peale}, S.~J. 1977, in IAU Colloq. 28: Planetary Satellites, ed. J.~A.
  {Burns}, 87--111

\bibitem[{{Pepe} {et~al.}(2011){Pepe}, {Lovis}, {S{\'e}gransan}, {Benz},
  {Bouchy}, {Dumusque}, {Mayor}, {Queloz}, {Santos}, \& {Udry}}]{pepe11}
{Pepe}, F., {et~al.} 2011, \aap, 534, A58

\bibitem[{{Pilbratt} {et~al.}(2010){Pilbratt}, {Riedinger}, {Passvogel},
  {Crone}, {Doyle}, {Gageur}, {Heras}, {Jewell}, {Metcalfe}, {Ott}, \&
  {Schmidt}}]{pilbratt2010}
{Pilbratt}, G.~L., {et~al.} 2010, \aap, 518, L1

\bibitem[{{Reid} {et~al.}(2002){Reid}, {Gizis}, \& {Hawley}}]{reid2002}
{Reid}, I.~N., {Gizis}, J.~E., \& {Hawley}, S.~L. 2002, \aj, 124, 2721

\bibitem[{{Reiners} {et~al.}(2010){Reiners}, {Bean}, {Huber}, {Dreizler},
  {Seifahrt}, \& {Czesla}}]{reiners2010}
{Reiners}, A., {Bean}, J.~L., {Huber}, K.~F., {Dreizler}, S., {Seifahrt}, A.,
  \& {Czesla}, S. 2010, \apj, 710, 432

\bibitem[{{Rivera} {et~al.}(2005){Rivera}, {Lissauer}, {Butler}, {Marcy},
  {Vogt}, {Fischer}, {Brown}, {Laughlin}, \& {Henry}}]{rivera2005}
{Rivera}, E.~J., {et~al.} 2005, \apj, 634, 625

\bibitem[{{Scargle}(1982)}]{scargle1982}
{Scargle}, J.~D. 1982, \apj, 263, 835

\bibitem[{{Udry} {et~al.}(2007){Udry}, {Bonfils}, {Delfosse}, {Forveille},
  {Mayor}, {Perrier}, {Bouchy}, {Lovis}, {Pepe}, {Queloz}, \&
  {Bertaux}}]{udry07}
{Udry}, S., {et~al.} 2007, \aap, 469, L43

\bibitem[{{Valenti} {et~al.}(1995){Valenti}, {Butler}, \&
  {Marcy}}]{valenti1995}
{Valenti}, J.~A., {Butler}, R.~P., \& {Marcy}, G.~W. 1995, \pasp, 107, 966

\bibitem[{{Wang} \& {Ge}(2011)}]{wang2012}
{Wang}, J., \& {Ge}, J. 2011, ArXiv e-prints

\bibitem[{{Wright} {et~al.}(2011){Wright}, {Fakhouri}, {Marcy}, {Han}, {Feng},
  {Johnson}, {Howard}, {Fischer}, {Valenti}, {Anderson}, \&
  {Piskunov}}]{wright2011}
{Wright}, J.~T., {et~al.} 2011, \pasp, 123, 412

\end{thebibliography}
\end{document}